# Positive streamers in ambient air and a N$_2$:O$_2$-mixture (99.8 : 0.2)

T.M.P. Briels, E.M. van Veldhuizen, U. Ebert

*Abstract* – **Photographs show distinct differences between positive streamers in air or in a nitrogen-oxygen mixture (0.2% O$_2$). The streamers in the mixture branch more frequently, but the branches also extinguish more easily. Probably related to that, the streamers in the mixture propagate more in a zigzag manner while they are straighter in air. Furthermore, streamers in the mixture can become longer; they are thinner and more intense.**

Streamers are narrow rapidly growing ionized channels. They can be created when a high voltage is applied to a nonconducting medium [1]. They are used for example in gas and water cleaning [1,2]. In nature they are observed as so-called sprite discharges in the atmosphere at 40 to 90 km altitude [3]. Streamers are often investigated in ambient air since this is the most commonly used gas in applications, experiments and nature. However, air is a compound gas in which many processes can occur. For understanding the physical mechanisms it is useful to perform experiments in simple gases as well. N$_2$ is a good candidate because it is the main component of air and it is a simple single molecular gas. Note, however, that small impurity concentrations can be essential and can never be fully suppressed. Therefore, we here present experiments where the O$_2$-concentration in N$_2$ is varied by a factor of 100, i.e. the gases used are ambient air and a nitrogen-oxygen mixture with 99.8% N$_2$ and 0.2% O$_2$ (hereafter abbreviated to N$_2$:O$_2$). This mixture is taken from a bottle. Comparable studies [4,5] report different pictures.

High-voltage pulses are created using a switched capacitor supply [6,7]. The voltage pulse is intentionally given a long rise time so that only thin streamers are created [6,7]. Photographs are taken with a 4QuikE intensified CCD-camera from Stanford Computer Optics.

Figure 1 shows positive streamers in a 160 mm point-plane gap in N$_2$:O$_2$ and air at 400 mbar and 30 kV. The discharge starts at the needle tip (top of photographs, indicated by 0 mm) and propagates towards the plate (bottom, 160 mm). The discharge in air is clearly not fractal. The discharge in N$_2$:O$_2$ forms many more branches and zigzags than the one in air. Furthermore, the many side branches in N$_2$:O$_2$ die out after a much shorter distance than in air. The discharge in N$_2$:O$_2$ branches roughly every 7.5 ± 2.5 mm while in air it branches roughly every 10 ± 4 mm. The discharge in N$_2$:O$_2$ is more intense, therefore the intensity level of the figures was reduced relatively to those in air. Streamers in N$_2$:O$_2$ are thinner, show a better contrast between in- and out-of-focus, and have diffuser tips. Other observations show that streamers in N$_2$:O$_2$ propagate further in space and have longer current pulse durations than streamers in air under similar conditions [7].

The difference in current pulse duration can be explained by the electronegative character of O$_2$ in air which attaches the electrons that are necessary to maintain a discharge. The discharge in air therefore will die out sooner because of an electron shortage. The difference in branching statistics can most likely be ascribed to photoionization. Also simulations [8,9] in a fluid model show that branching can be delayed by photoionization. However, this should be reinvestigated in a particle model with its inherent particle density fluctuations [10]. The average electron energy in the streamer head is deduced from the electric field [11] and the field is obtained via the ratio of the spectral lines of N$_2$ [12]. The obtained results show higher average electron energy for N$_2$ which can explain the observed difference in intensity [7]. The experiments will be discussed in more detail in [13].

Manuscript received X XXX 2007; revised X XXX 2007. The paper has been accepted for IEEE Trans. Plasma Sci.and is scheduled to appear in June 2008.
T.M.P. Briels, E.M. van Veldhuizen and U. Ebert are with Dept. Appl. Phys., Eindhoven Univ. Techn., P.O. Box 513, 5600 MB Eindhoven, The Netherlands (e-mail: e.m.v.veldhuizen@tue.nl). U. Ebert is also with Centrum Wiskunde & Informatica (CWI) Amsterdam, The Netherlands.

This research was supported by STW, project 06501, and by The Netherland's Organisation for Scientific Research (NWO).

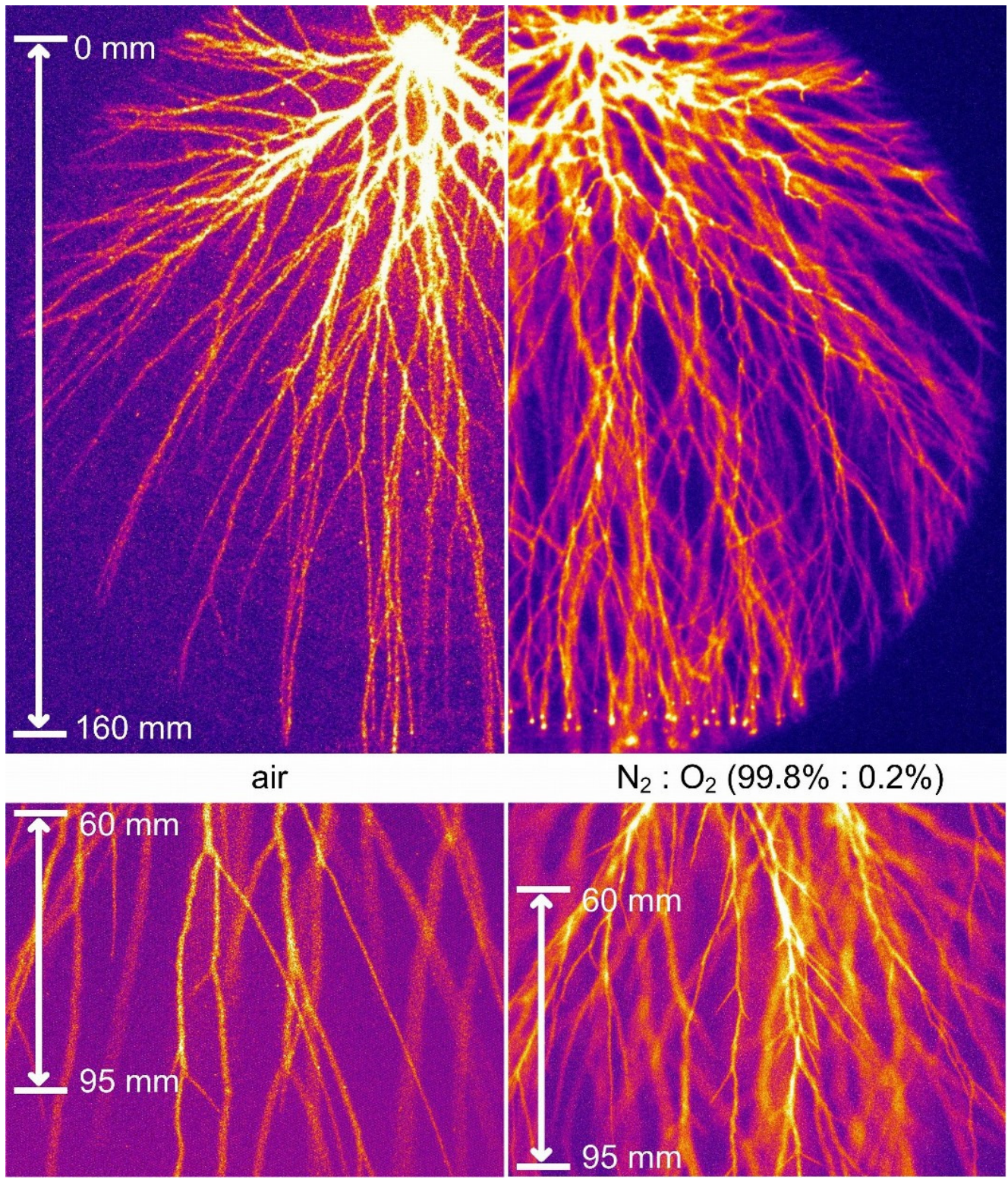

Figure 1. Positive streamers at 400 mbar and 30 kV in a 160 mm gap in air (left) and 99.8% $N_2$:0.2% $O_2$ (right). The top photographs show the complete discharge. The dark ring around the photographs is the edge of the viewing port of the setup. The gate delay and gate width (cf. [6]) are 0 μs and 70 μs in air and 0 μs and 5 μs in $N_2$:$O_2$, respectively. The bottom photographs zoom into the middle region of the 160 mm gap at a position of 60 to 95 mm from the anode tip. The gate delay and gate width are 0 μs and 4 μs in air and 1 μs and 3 μs in $N_2$:$O_2$, respectively. Note that the photographs cannot be compared in intensity since the streamers in $N_2$:$O_2$ are much more intense and thus recorded with different camera settings.